\documentclass[traditabstract]{aa}  

\usepackage{amsmath}
\usepackage{amssymb}
\usepackage{natbib}
\usepackage{graphicx}
\usepackage{comment}
\usepackage{txfonts}
\usepackage{ctable}
\usepackage{hyperref}
\usepackage{xspace}
\usepackage{epstopdf, epsfig}
\usepackage{tabularx}
\usepackage{threeparttable}
\usepackage{caption}
\usepackage{subcaption}
\usepackage{hyperref}

\def\pwn{HESS\,J1303$-$631}
\def\pulsar{PSR\,J1301$-$6305}

\def\snrpmn{PMN\,J1303$-$6259}
\def\snr{G304.4$-$0.2}

\def\gammaray{$\gamma$-ray}
\def\gammarays{$\gamma$-rays}

\def\hess{H.E.S.S.}
\def\chandra{\textit{Chandra}}
\def\xmm{\textit{XMM-Newton}}
\def\asca{\textit{ASCA}}
\def\rosat{\textit{ROSAT}}
\def\swift{\textit{Swift}}
\def\fermi{\textit{Fermi}-LAT}
\def\atca{ATCA}
\def\deg{$^{\circ}$}

\newcommand{\rr}[1]{#1}
\newcommand{\rrr}[1]{#1}
\newcommand{\R}[1]{#1}
\newcommand{\RR}[1]{#1}
\newcommand{\RRR}[1]{#1}
\newcommand{\RRRR}[1]{#1}

\begin{document}

%\begin{frontmatter}



\title{Radio Observations of the \rrr{Region around the} Pulsar Wind Nebula \pwn\ with \atca}


\author{Iurii Sushch \inst{1,2,3}
\and Igor Oya \inst{2}
\and Ullrich Schwanke \inst{4}
\and Simon Johnston \inst{5}
\and Matthew L. Dalton \inst{6}}
\institute{Centre for Space Research, North-West University, Potchefstroom 2520, South Africa
  \and
  DESY, D-15738 Zeuthen, Germany  
  \and
  Astronomical Observatory of Ivan Franko National University of L'viv, vul. Kyryla i Methodia, 8, L'viv 79005, Ukraine
  \and 
  Institut f\"{u}r Physik, Humboldt-Universit\"{a}t zu Berlin, Newtonstr. 15,
  D 12489 Berlin, Germany
  \and 
  Australia Telescope National Facility, CSIRO, PO Box 76, Epping, NSW 1710, Australia
  \and 
  Active Space Technologies GmbH, Carl Scheele Strasse 14, 12489 Berlin, Germany
}
\date{Received ?, 2015; accepted ?, ?}

%\address[ohio]{Astrophysical Institute, Department of Physics and Astronomy, Ohio University, Athens, OH 45701, USA}
\abstract{
Radio observations of \rr{the region surrounding} \pulsar\ at $5.5$\,GHz and $7.5$\,GHz 
were conducted with \atca\ on September 5th, 2013. They were dedicated to the search 
of the radio counterpart of the evolved pulsar wind nebula \pwn, \rr{detected in X-rays and GeV-TeV \gammarays}. 
The collected data do not reveal any significant extended emission associated with \pulsar. In addition, archival 
$1.384$\,GHz and $2.368$\,GHz data do not show any evidence for a radio 
counterpart of \pwn. Archival $1.384$\,GHz observations reveal a \rr{detection} of \RRR{an extended}
structure \R{centred} at an angular distance of \R{$19^\prime$} from the pulsar. 
This \RRR{extended} structure might be a Supernova remnant (SNR) and a potential birth place of \pulsar. 
The implications of the lack of radio counterpart of \pwn\ on the understanding of the 
nature of the PWN are discussed.
}

\keywords{pulsars: individual: \pulsar; stars: winds, outflows; radio continuum: stars; radiation mechanisms: non-thermal}

%\begin{keyword}



%\end{keyword}

%\end{frontmatter}


\authorrunning{Sushch et al.}
\titlerunning{Radio Observations of the Region around the Pulsar Wind Nebula \pwn\ with \atca}
\maketitle

\section{Introduction}
\pwn\ is one of the most prominent examples of the so-called very
  high energy (VHE; $E>100$\,GeV) \gammaray\ \rr{``dark''} sources, those 
which were detected in the VHE band but do not have counterparts
at other energy bands. It was discovered in 2005
\citep{2005A&A...439.1013A} but the nature of the source was unclear
until 2012, when a detailed study of the energy-dependent morphology
provided evidence of the association with the pulsar \pulsar\ 
\citep{2012A&A...548A..46H}. With the increase of the energy threshold
a very extended emission region  ($\sim 0.4$\deg$\times 0.3$\deg\ at the 
$[0.84 - 2]$ TeV band)  of VHE \gammarays\ \rr{``shrinks''} towards the
position of the pulsar at $E> 10$\,TeV. \rr{While at lower energies the peak 
position of the extended emission region is significantly offset from the 
position of the pulsar, at energies above 10\,TeV the pulsar is coincident with 
the peak of the \gammaray\ emission region \citep{2012A&A...548A..46H}.} 
Such an energy-dependent morphology is expected 
%in a leptonic scenario 
for ancient pulsar wind nebulae \RR{\citep[PWNe;][and references therein]{2012A&A...548A..46H}}. 
%% , which feature several
%% populations of relativistic electrons generated by the pulsar.  
Young electrons located close to the pulsar are not cooled yet and, thus,
very energetic. These energetic electrons generate
the VHE emission around the pulsar via inverse Compton (IC)
scattering on the cosmic microwave background photons (CMB). 
Older, cooled down, lower energy electrons might be spread farther 
away from the pulsar for several reasons (e.g. proper motion of the 
pulsar which causes that older particles are left behind and/or particle 
diffusion) but they can still produce \gammarays\ via IC scattering, however 
at lower energies than the young electrons.

The association of \pwn\ with the pulsar is further supported 
by the detection of its X-ray counterpart with \xmm\ 
\citep{2012A&A...548A..46H}. The size of the X-ray PWN is much 
smaller than the size of the VHE source, extending $2^\prime-3^\prime$ from the 
pulsar position towards the centre of the VHE \gammaray\ emission region. The
much smaller size of the X-ray emitting region can be explained by an effective 
synchrotron cooling of older electrons to energies too low to generate 
synchrotron emission in the X-ray energy range and/or due to the decreasing
magnetic field strength in the PWN with time 
\rr{\citep[see e.g.][]{2009arXiv0906.2644D, 2013ApJ...773..139V}}. The tail-like extension of the 
X-ray source might be an indication of the proper motion direction of the 
pulsar triggering speculations about its possible birth-place 
\citep{2012A&A...548A..46H}. 

An analysis of archival data from the 
Parkes-MIT-NRAO (PMN) survey at 4.85 GHz 
\citep{1993AJ....106.1095C} revealed also 
a hint of radio emission at the pulsar position with size 
comparable to the X-ray emission region \RR{\citep{2012A&A...548A..46H}}. Data analysis showed a 
$\sim3\sigma$ feature with a peak flux \R{density} of \rrr{$30$ mJy/beam} which is at the 
detection limit of the survey \RR{and was considered as an upper limit}. This hint of a 
radio counterpart of \pwn\ triggered new dedicated observations with the 
Australian Telescope Compact Array (\atca), which were conducted in 
September 2013. Results of these observations are presented in this paper.

Recently, the counterpart of \pwn\ was finally detected at GeV
energies with \fermi\ \citep{2013ApJ...773...77A}. \rr{The morphology of the source 
is consistent with a Gaussian of width $0.45^{\circ}$.} The source is
contaminated by the emission of the nearby Supernova remnant (SNR)
Kes\,17 \rr{(G$304.6+0.1$)}, but it is clearly seen above $31$\,GeV. The emission region of the 
GeV counterpart of \pwn\ is as expected larger than the TeV source, but
the morphology of the emission region is very similar and features an
extension in the same direction as the TeV source. \rr{It should be noted, however, 
that the size of the GeV source might be slightly overestimated due to the contamination 
from Kes\,17. Kes\,17 is the closest known SNR to the pulsar \pulsar\ located at the angular distance of 
$37^{\prime}$ \citep{2011ApJ...740L..12W, 2013ApJ...777..148G}. Assuming a distance to the pulsar of $6.6$\,kpc 
\citep{2012A&A...548A..46H}, this corresponds to the projected distance between the pulsar and the 
SNR of 71\,pc. This large distance makes the association of the SNR with the pulsar very unlikely as it would require an 
unrealistically high pulsar velocity of $\sim6,000$\,km/s for the 
characteristic age of the pulsar of 11 \RRR{kyr} \citep{2005AJ....129.1993M}.}

%\section{Observations and Data Analysis Results}
\RR{\section{Observations}}

The \atca\ observations of the \rrr{region surrounding} \pulsar\ %PSR J1301-6305 
were conducted on September 5th, 2013.  Observations were performed \rr{using the CABB receiver} 
with the $1.5$A configuration of the array \rrr{(minimum and maximum baselines of 153\,m and 3000\,m, respectively)}
%% \rr{with a bandwidth of 2048 MHz over 
%% 2049 channels with one band centered at $5.5$ and the other at $7.5$\,GHz.} 
at $5.5$ and $7.5$\,GHz frequencies and 
%\rr{The observations were centered} 
centred at $\alpha =
13^{\mathrm{h}}02^{\mathrm{m}}10.00^{\mathrm{s}}$, $\delta =
-63^{\circ}05^{\prime}34.8^{\prime\prime}$ (J2000.0), \rr{at the angular distance of about $3^\prime$ from the pulsar position.} 
\rrr{The array configuration was chosen in order to match the resolution of the \xmm\ observations of $\sim 4^{\prime\prime}$, 
while at the same time remain sensitive to structures comparable to the size of the X-ray PWN of $\sim2^{\prime}$. \R{However, the maximum 
angular scales to which the observations are sensitive are slightly smaller than the size of the X-ray PWN, namely $\sim1.7^{\prime}$ at 5.5\,GHz and $\sim1.3^{\prime}$ at 7.5\,GHz, respectively.}%\footnote{\R{Values were estimated from Table 1.5 of the ATCA Users Guide, Revision 1.5.}} 
}

%\rr{
  The observations were carried out in two modes: {\it CFB 1M (no zooms)} - a bandwidth of 2 GHz with 2048 1-MHz channels in each intermediate frequency 
  (IF) band and {\it pulsar binning} - the same but with the addition of pulsar binning according to the provided ephemerids. The on-source scan time was \R{656.5 min.}
 % $2626.7$ \rrr{min}.} 
{\it Pulsar binning} mode was used in observations in order to be able to correctly subtract the pulsar contribution \R{to the total emission in order to determine the intrinsic
    emission from \pwn}. However, since no significant \R{emission corresponding to HESSJ1303-631}
  %\rrr{extended} emission
  was detected (see Section \ref{data_analysis}), \R{the subtraction of the data taken in pulsar binning mode was not performed and thus these data were not used in this study.}
%  the data taken in {\it pulsar binning} mode were not used in this study.}    
Primary \rr{(flux density)} and secondary \rr{(phase)} calibrators
were J$1934-638$ and J$1352-63$ respectively. \rrr{The flux density of J$1934-638$ is $5.00$\,Jy at $5.5$\,GHz and $2.97$\,Jy at $7.5$\,GHz.} 
\rr{The phase calibrator was observed every \rrr{$\sim30$\,min}. The observation recorded all four linear polarization modes.} 
Details of the collected data are listed in Table \ref{archive_data}.

In this paper we also considered archival \atca\ data obtained during 
observations of \pwn\, centred at $\alpha = 13^{\mathrm{h}}03^{\mathrm{m}}0.400^{\mathrm{s}}$, $\delta = -63^{\circ}11^{\prime}11.55^{\prime\prime} $
\rr{(the position of the peak VHE \gammaray\ emission)}, 
and performed in the $1.384$\,GHz and $2.368$\,GHz bands.
The archival data used in the analysis, taken as part of the 
Reinfrank\,et\,al. project C1557, are presented in Table \ref{archive_data}. 
Only the archival data taken with all 6 antennas and with observational time 
longer than 100 \rrr{min} were used in the analysis. \rrr{The source J$1934-638$ was used as a 
primary calibrator with flux densities of $14.95$\,Jy at $1.384$\,GHz and $11.59$\,Jy at $2.368$\,GHz. The sources J$1421-490$ and J$1329-665$ were used 
for phase calibration. The maximum angular scale to which observations are sensitive is \R{$\sim18^{\prime}$} at $1.384$\,GHz and \R{$\sim11^{\prime}$} at $2.368$\,GHz.}

\begin{table*}
\centering
\caption{Details of the \atca\ data of the \pwn\ \rrr{region} analysed in
  this paper}
\label{archive_data}
%\begin{tabular}{c c c c c c c c c}
\begin{tabular}{l l l l l l l l l}
\hline
\hline
\\
Date & Right Ascention& Declination& Time  & Array& Frequencies & \rr{Bandwidth} & \rr{Primary} & \rr{Secondary}\\
     &                &            & [min] &      & [MHz]       & \rr{[MHz]} & \rr{calibrator} & \rr{calibrator}\\
\hline
\\
2013-Sep-05& $13^\mathrm{h}2^\mathrm{m}10.00^\mathrm{s}$& $-63^{\circ}5^\prime34.8^{\prime\prime}$&\R{656.5}& 1.5A& 5500, 7500& \rr{2048}& \rr{$1934-638$}& \rr{$1352-63$}\\
2006-Oct-25& $13^\mathrm{h}3^\mathrm{m}0.400^\mathrm{s}$& $-63^{\circ}11^\prime11.55^{\prime\prime}$&433.8&EW352&1384, 2368& \rr{128}& \rr{$1934-638$}& \rr{$1421-490$}\\
2007-Mar-13& $13^\mathrm{h}3^\mathrm{m}0.400^\mathrm{s}$& $-63^{\circ}11^\prime11.55^{\prime\prime}$&618.1&750D&1384, 2368& \rr{128}& \rr{$1934-638$}& \rr{$1329-665$}\\
2007-Apr-24& $13^\mathrm{h}3^\mathrm{m}0.400^\mathrm{s}$& $-63^{\circ}11^\prime11.55^{\prime\prime}$&651.8&1.5C&1384, 2368& \rr{128}& \rr{$1934-638$}& \rr{$1329-665$}\\	
\hline
\end{tabular}
\end{table*}

\begin{figure}
\centering
\resizebox{\hsize}{!}{\includegraphics{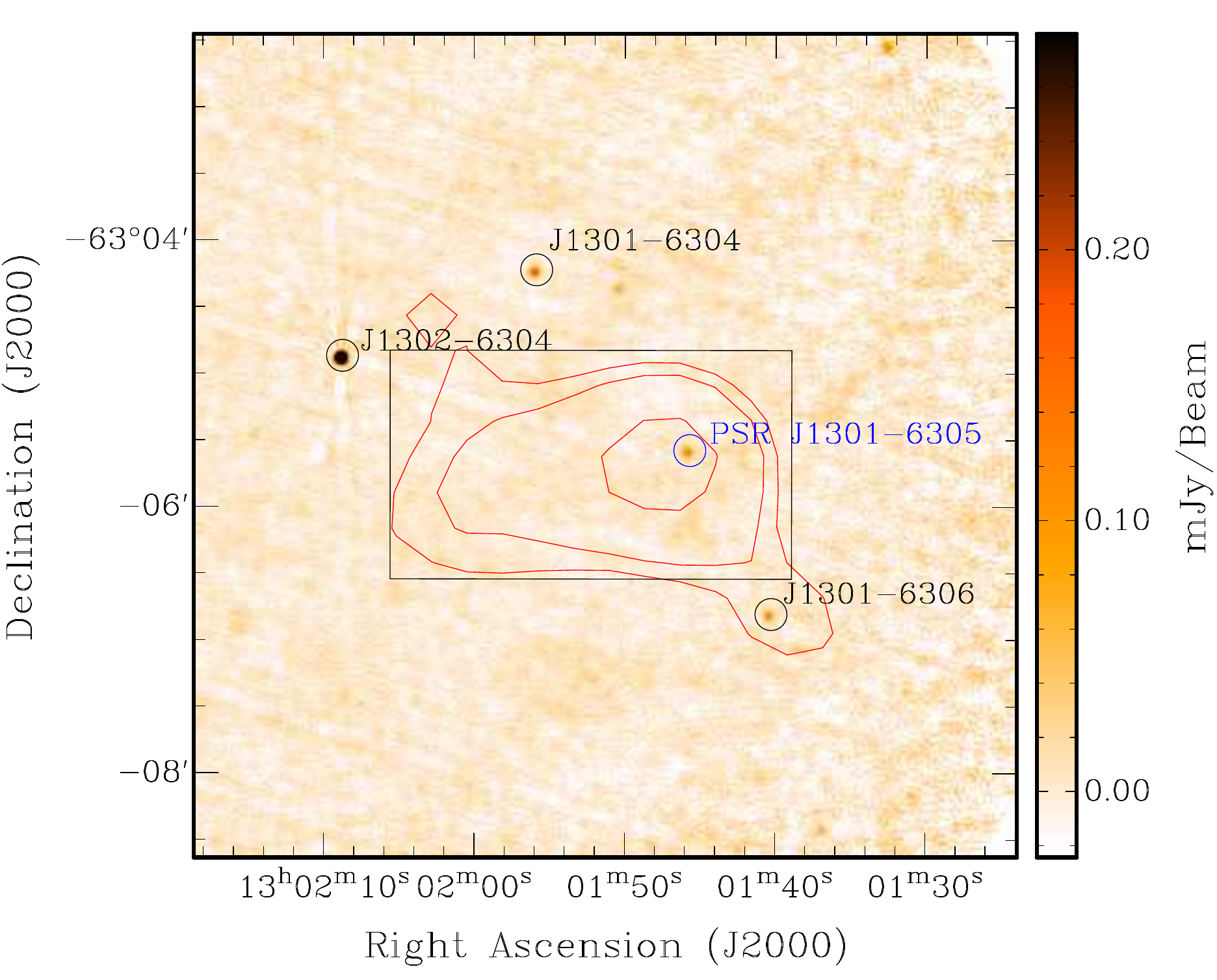}}
\caption{Radio map of the \pwn\ \rrr{region} at $5.5$\,GHz
  overlaid with 
%% significance contours from the  $1.384$ GHz band 
%% (from $2\sigma$ to $6\sigma$ with the interval of $2\sigma$ and from $10\sigma$ to $70\sigma$ with 
%% the interval of $20\sigma$). 
contours of the X-ray PWN (red) as detected with \xmm\ \citep{2012A&A...548A..46H}. \RRR{X-ray contours show the peak of the emission close to the pulsar with a trail expanding in the eastern direction.} The blue circle indicates the position of the pulsar PSR J1301-6305. The black 
box determines the region used for the flux \R{density} upper limit calculation. \rr{New point-like sources are indicated with index numbers. The synthesised beam is determined  by an ellipse with 
the major and minor axes of $3.79^{\prime\prime}$ and $3.65^{\prime\prime}$ respectively and the positional angle of $-7.6^{\circ}$.}}
\label{radiomap}
\end{figure}

\RR{\section{Data Analysis \rr{and Results}}}
\label{data_analysis}
\rr{The data reduction and image analysis was performed using the \texttt{miriad}
\citep{1995ASPC...77..433S} and \texttt{karma} \citep{1995ASPC...77..144G} packages.} The resulting clean 
primary beam corrected (restricted to the area of primary beam 
response above $30\%$, \rr{which corresponds to the radial distance of $5.8^{\prime}$}) image \rr{(Stokes I)} 
at $5.5$\,GHz is shown in Fig.\,\ref{radiomap}. 
The pulsar \pulsar\ is detected at the position $\alpha = 13^\mathrm{h}01^\mathrm{m}45.678^\mathrm{s} \pm
0.013^\mathrm{s}$, $\delta = -63^{\circ}05^\prime34.85^{\prime\prime}
\pm 0.20^{\prime\prime}$.  No significant extended emission
coincident with the pulsar position was detected. The fitted image
root mean square (RMS) noise is calculated using the \texttt{imsad} task at
the level of \rrr{$0.011$\,mJy/beam}. \rr{The synthesised beam is an ellipse with 
the major and minor axes of $3.79^{\prime\prime}$ and
$3.65^{\prime\prime}$ respectively and the positional angle of $-7.6^{\circ}$}.

There is no extended emission coincident with the pulsar position
detected at $7.5$\,GHz as well. The fitted image RMS noise,
calculated using the \texttt{imsad} task, is at the level of \rrr{$0.011$\,mJy/beam}. \rr{The major and minor axes of the beam are 
$3.06^{\prime\prime}$ and $2.90^{\prime\prime}$,
respectively, and the positional angle is $-6.2^{\circ}$}. 
%% \iocom{QUESTION: Not figure for this?} \is{Do we really need showing another "empty" map? It will look pretty much the same as 5.5 GHz one.}

\begin{figure*}[ht!]
  \centering
  %\begin{subfigure}[b]{0.5\textwidth}
%    \includegraphics[width=\linewidth]{1384_paper_1000_10_corrsour.eps}
%    \resizebox{\hsize}{!}{\includegraphics{1384_paper_1000_10_corrsour.eps}}
    \resizebox{\hsize}{!}{\includegraphics{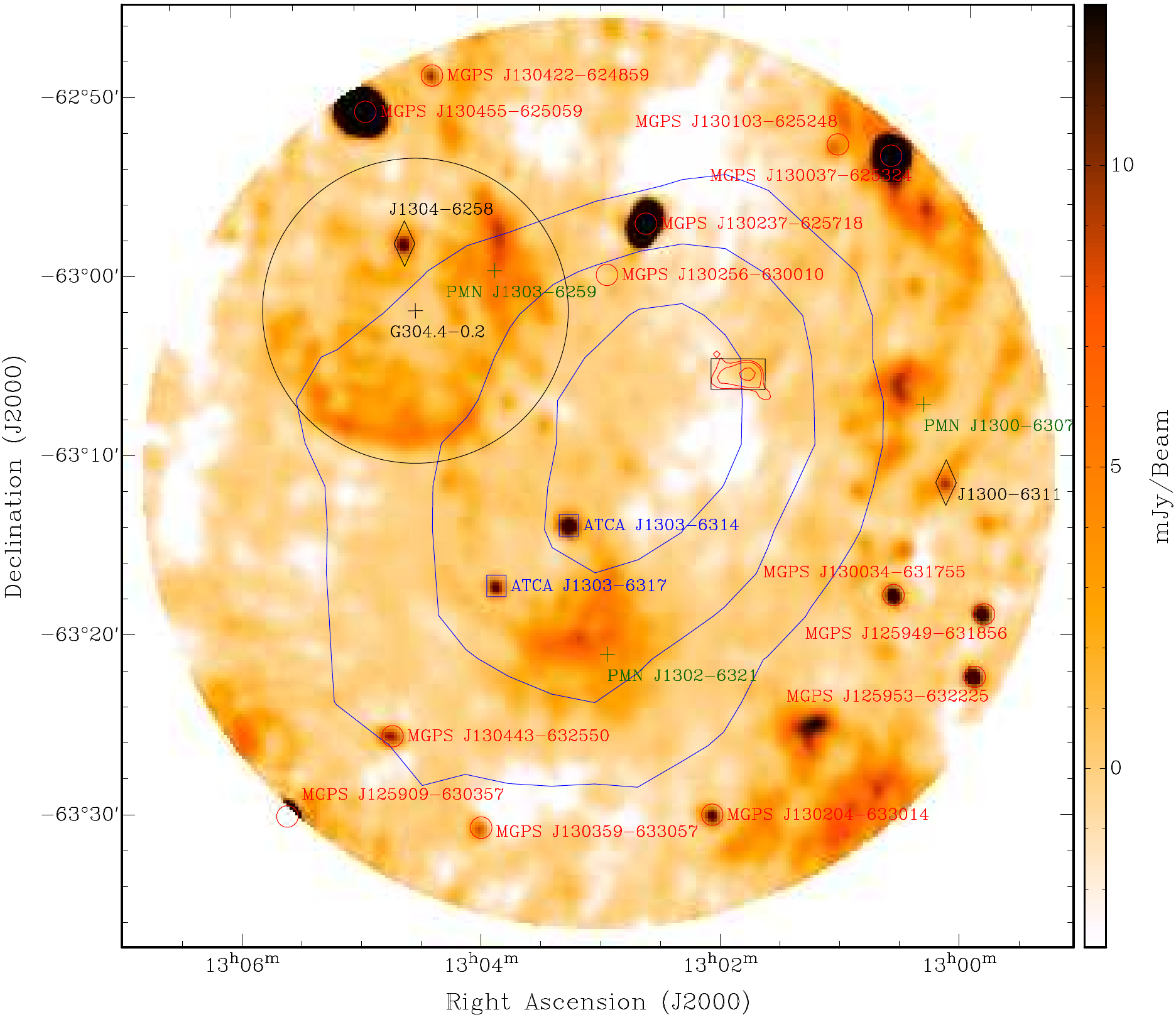}}
      %\resizebox{\hsize}{!}{\includegraphics{1384_cont1384_2_6+2_10_70+20.eps}}
  %\end{subfigure}
  %% \begin{subfigure}[b]{0.5\textwidth}
  %%       \includegraphics[width=\textwidth]{slice.eps}
  %% \end{subfigure}
    %% \caption{Top: Radio map of the \pwn\ \rrr{region} at $1.384$\,GHz. 
    %%   %% Contour levels represent $2\sigma$,$4\sigma, and $6\sigma$  $1.384$\,GHz emission. The red cross indicates the position of the pulsar PSR J1301-6305. The blue circle shows the result of the disk fit to the SNR candidate and the blue cross indicates the centre of the disk.
    %%   Red contours represent the X-ray PWN as detected with \xmm\ and blue contours represent \pwn\ as detected by \hess\ \citep{2012A&A...548A..46H}. The black circle 
    %%   indicates the shell-like structure \rr{and the black cross indicates its centre}. \rr{New point-like sources are indicated with index numbers. The synthesized beam is determined  by an ellipse with the major and minor axes of $51.3^{\prime\prime}$ and $43.8^{\prime\prime}$ respectively and the positional angle of $20.9^{\circ}$. Bottom: 1D radial profile of the shell-like structure, i.e. flux density along the black line shown on the top image. Red dashed line indicates a $3\,\sigma$ level.}}
\caption{Radio map of the \pwn\ \rrr{region} at $1.384$\,GHz. 
      %% Contour levels represent $2\sigma$,$4\sigma$, and $6\sigma$  $1.384$\,GHz emission. The red cross indicates the position of the pulsar PSR J1301-6305. The blue circle shows the result of the disk fit to the SNR candidate and the blue cross indicates the centre of the disk.
      Red contours represent the X-ray PWN as detected with \xmm\ and blue contours represent \pwn\ as detected by \hess\ \citep{2012A&A...548A..46H}. The black circle 
      indicates the SNR candidate \snr\ \rr{and the black cross indicates its centre}. \rrr{MGPS-2 catalogue sources \citep{2007MNRAS.382..382M} are shown as red circles, PMN survey sources \citep{1993AJ....106.1095C}, excluding those which are coincident with MGPS-2 sources, are shown as green crosses and two ATCA sources \citep{2015ApJS..217....4S} are shown as blue squares. The two new point-like radio sources reported in this paper are marked with black diamonds.} 
%\rr{New point-like sources are indicated with index numbers.} 
\rr{The synthesised beam is determined  by an ellipse with the major and minor axes of $51.3^{\prime\prime}$ and $43.8^{\prime\prime}$ respectively and the positional angle of $20.9^{\circ}$.}}
    \label{radiomap1384}
  \end{figure*}

The analysis of the archival data at $1.384$\,GHz and $2.368$\,GHz 
which combine all the observations listed in Table \ref{archive_data} 
also does not reveal any significant emission coincident with the pulsar. 
%% The $1.384$\,GHz (Fig.\,\ref{radiomap1384}) and $2.368$\,GHz flux maps which combine all the archival data 
%% listed in Table \ref{archive_data} also do not reveal any significant emission 
%% coincident with the pulsar. 
The observations at $1.384$\,GHz, however, reveal a \rr{detection} of a shell-like structure to
the east of the pulsar position which might 
potentially be an SNR \RR{(see discussion in Section \ref{snr})}. Figure \ref{radiomap1384} shows the cleaned and 
primary beam corrected (restricted to the area of primary beam 
response above $20\%$, \rr{which corresponds to the radial distance of $25.3^{\prime}$}) image at $1.384$\,GHz. The \rrr{SNR candidate \snr} is 
positioned \rr{within} the black circle. \rr{The centre of the structure is at $\alpha = 13^\mathrm{h}04^\mathrm{m}31.1^\mathrm{s}$, 
$\delta = -63^{\circ}02^\prime08^{\prime\prime}$.} The fitted image RMS noise is estimated
to be \rrr{$0.697$\,mJy/beam}. \rr{The major and minor axes of the
beam are \rr{$51.3^{\prime\prime}$} and \rr{$43.8^{\prime\prime}$}, respectively, and the positional angle is \rr{$20.9^\circ$}.} 
The brightest parts of the \rrr{SNR candidate}
reach the significance of $13\,\sigma$. 
\rrr{It is difficult to draw any conclusions about a possible emission from the SNR candidate 
  at $2.368$\,GHz, as these observations are less sensitive to large scale structures, and only a fraction 
of the SNR candidate is located within the primary beam and the image is distorted 
by artefacts produced by the strong source MGPS\,J$130237-625718$.}

\begin{table*}
\centering
\caption{New point-like radio sources detected in these observations}
\label{newsources}
\begin{threeparttable}
\rrr{
%\begin{tabular}{c c c c c c c c c}
\begin{tabular}{l l l l l l l l}
\hline
\hline
\\
Identifier & Right Ascention& Declination& $F_{5.5\,\mathrm{GHz}}$& $F_{7.5\,\mathrm{GHz}}$& $F_{1.384\,\mathrm{GHz}}$& $F_{2.368\,\mathrm{GHz}}$  & X-ray counterparts\tnote{a} \\
           &                &            &  [$\mu$Jy]        & [$\mu$Jy]         &[mJy]           & [mJy]               &      \\
\hline
\\

J1301-6306 & $13^\mathrm{h}01^\mathrm{m}40.27^\mathrm{s}$& $-63^{\circ}06^\prime48.6^{\prime\prime}$&127.2& 118.3 & ... & ... & 3XMM J130138.2-630654\tnote{b} \\
J1302-6304 & $13^\mathrm{h}02^\mathrm{m}8.74^\mathrm{s}$& $-63^{\circ}04^\prime52.1^{\prime\prime}$ &1448.0 &  1172.0& ... & ... & none\\
%3     & 1301-6304 & $13^\mathrm{h}01^\mathrm{m}50.25^\mathrm{s}$& $-63^{\circ}04^\prime21.0^{\prime\prime}$&80.5&   65.5& ... & ... & none\\
J1301-6304 & $13^\mathrm{h}01^\mathrm{m}55.86^\mathrm{s}$& $-63^{\circ}04^\prime13.5^{\prime\prime}$&191.2&  119.1& ... & ... & none\\
J1300-6311 & $13^\mathrm{h}0^\mathrm{m}9.23^\mathrm{s}$& $-63^{\circ}11^\prime37.8^{\prime\prime}$&...&  ...& 11.7 & ... & 3XMM J130006.2-631207\tnote{b}\\
%% 1300-6311 :  13:00:09.06 -63:11:40.8  9.139E-03  9.139E-03  2.933E+01  2.649E+01 -7.860E+01 C F
J1304-6258 & $13^\mathrm{h}4^\mathrm{m}36.23^\mathrm{s}$& $-62^{\circ}58^\prime22.7^{\prime\prime}$&...&  ...& 11.3 & 5.7 & none\\
%% 1304-6258 :  13:04:36.46 -62:58:25.7  1.347E-02  1.347E-02  2.933E+01  2.649E+01 -7.860E+01 F F
%% 14    & 1301-6252 & $13^\mathrm{h}1^\mathrm{m}5.08^\mathrm{s}$& $-62^{\circ}52^\prime56.1^{\prime\prime}$&...&  ...& 8.0 & ... & none\\
%% %% 1301-6252 :  13:01:05.47 -62:52:57.6  8.228E-03  8.228E-03  2.933E+01  2.649E+01 -7.860E+01 C F
%% 15    & 1304-6248 & $13^\mathrm{h}4^\mathrm{m}22.50^\mathrm{s}$& $-62^{\circ}48^\prime58.0^{\prime\prime}$&...&  ...& 10.3 & ... & none\\
%% %% 1304-6248 :  13:04:22.35 -62:48:59.4  1.052E-02  1.052E-02  2.933E+01  2.649E+01 -7.860E+01 C F

\hline
\end{tabular}
%% \begin{tablenotes}
%% \item[a] within the radius of $10^{\prime\prime}$ for the radio map at 5.5\,GHz and within the radius of $1^\prime$ for the radio map at 1.384\,GHz
%% \item[b] a hint of a double-component morphology
%% \item[c] \citet{2012MNRAS.424.2442S}
%% \item[d] a hint of extended emission at 2.368\,GHz
%% \end{tablenotes}
\begin{tablenotes}
\item[a] \R{There are also multiple infrared and/or optical sources that are consistent with the position of these radio sources.}
\item[b] \citet{2016A&A...590A...1R}
\end{tablenotes}
}
\end{threeparttable}
\end{table*}

\RR{\subsection{Other sources in the observed region}}

\rrr{\R{Most of the} sources detected in the field at $1.384$\,GHz (Fig.\,\ref{radiomap1384}), both compact and extended, have counterparts at other radio frequencies 
\citep{2007MNRAS.382..382M, 1993AJ....106.1095C, 2015ApJS..217....4S}. However, most of these sources are not classified, except 
MGPS J$130422-624859$ (G$304.5-0.1$) which is identified as a HII region \citep{2002MNRAS.335..114M} and MGPS J$130037-625324$ which 
is coincident with an infrared bubble \citep{2012MNRAS.424.2442S}. Extended emission at the south-western edge of the field is most probably 
related to the very extended source PMN\,J$1259-6337$ \citep{1994ApJS...91..111W}.}

\begin{figure}[h!]
  \centering
  \resizebox{\hsize}{!}{\includegraphics{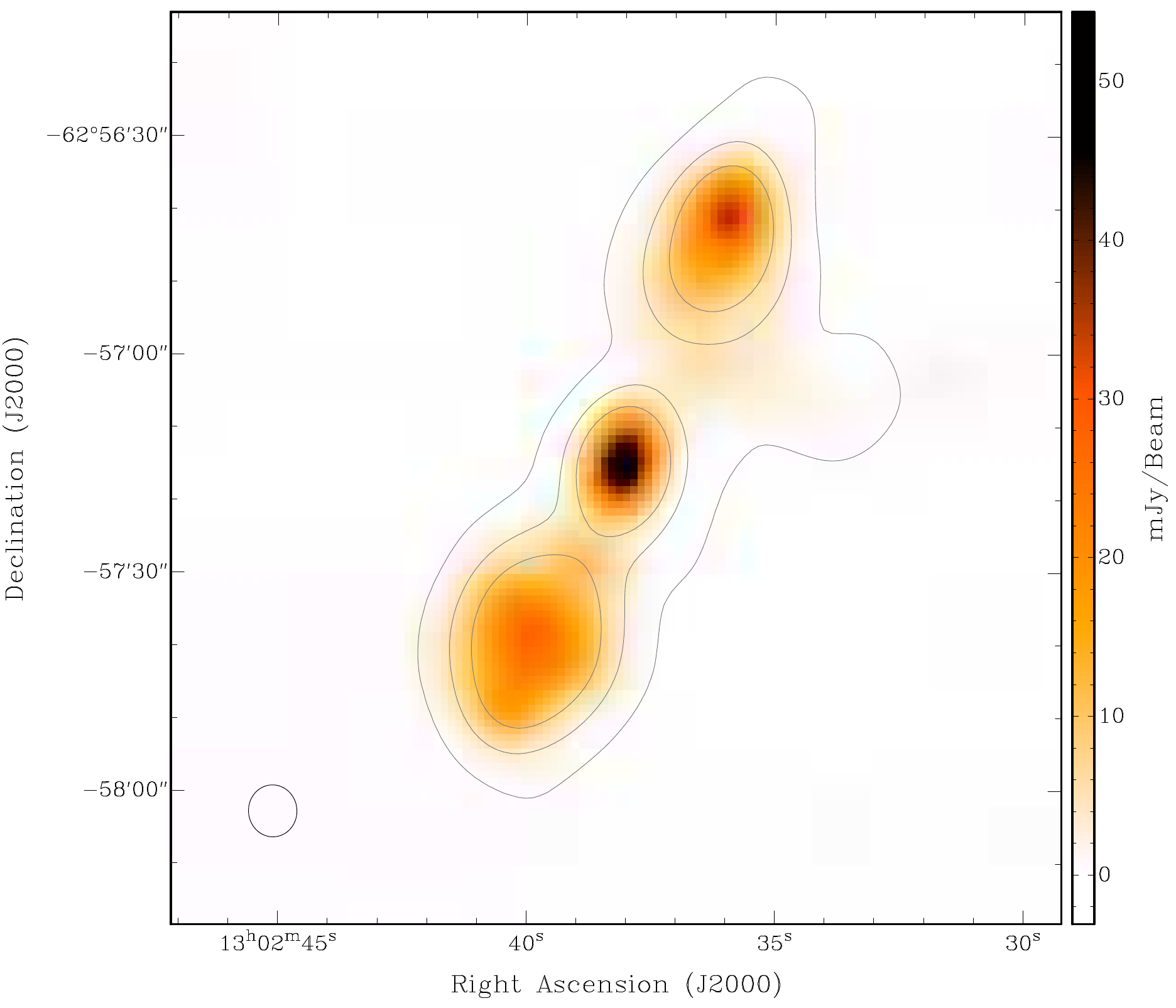}}
  %mgps2: scale -0.002 0.017
  % pmn: scale -0.05 0.5
  %1.384: scale -0.003 0.013

  %% \begin{subfigure}[b]{0.32\textwidth}
  %%   \includegraphics[width=\textwidth]{/home/lapuh/Nauka/PWN/J1303/1384_snrfov.pdf}
  %% \end{subfigure}
  %% \begin{subfigure}[b]{0.32\textwidth}
  %%   \includegraphics[width=\textwidth]{/home/lapuh/Nauka/PWN/J1303/pmn_snrfov_nocb.pdf}
  %% \end{subfigure}
  %% \begin{subfigure}[b]{0.32\textwidth}
  %%   \includegraphics[width=\textwidth]{/home/lapuh/Nauka/PWN/J1303/mgps2_snrfov.pdf}
  %% \end{subfigure}
  \caption{\rrr{Radio map of MGPS\,J$130237-625718$ at $2.368$\,GHz overlaid with contours indicating the significance of the emission at $1.384$\,GHz \R{at the level of $10$, $40$ and $80$ times the RMS noise}. The synthesised beam is shown with a black ellipse in the left bottom corner.}}
  \label{agn}
\end{figure}

\rrr{Observations at $1.384$\,GHz and $2.368$\,GHz might shed some light on the nature of the unidentified source MGPS\,J$130237-625718$, which is 
also visible at other radio wavelengths (e.g. it is detected in the PMN survey). Both $1.384$\,GHz and $2.368$\,GHz \R{data} reveal a complex morphology with 
a central source and two lobes,
%(Fig.\,\ref{agn}),
strongly 
suggesting that MGPS\,J$130237-625718$ is an active galactic nucleus (AGN).
\R{Figure\,\ref{agn} shows the map at $2.368$\,GHz overlaid with contours indicating
significant emission from the source at $1.384$\,GHz.}
Only the observations with the best angular resolution (2007-Apr-24) 
were used in this analysis.
\R{The fitted image RMS noise is $0.448$\,mJy/beam at $1.384$\,GHz (beam size: major and minor axes of $10.98^{\prime\prime}$ and $9.78^{\prime\prime}$ respectively with
  the positional angle of $1.7^\circ$) and $0.322$\,mJy/beam at $2.368$\,GHz (beam size: major and minor axes of $7.15^{\prime\prime}$ and $6.65^{\prime\prime}$ respectively with
  the positional angle of $3.7^\circ$).}
A fit to the observed emission at $1.384$\,GHz and $2.368$\,GHz was done
using the \texttt{imfit} task with three Gaussian components. The integrated flux \R{density} from the central component is $136.8 \pm 5.0$\,mJy and $80.8\pm3.7$\,mJy 
at $1.384$\,GHz and $2.368$\,GHz, respectively. The flux \R{densities} from the southern and northern components are $260.1\pm7.3$\,mJy and $192.5\pm6.1$\,mJy at $1.384$\,GHz 
and $156.1\pm7.4$\,mJy and $99.9\pm5.5$\,mJy at $2.368$\,GHz. Unfortunately, the source is outside the primary beam for observations at $5.5$\,GHz and $7.5$\,GHz 
with much better angular resolution.}

\rrr{In the \atca\ data presented here, 5 new point-like sources were detected in the region of \pwn\ (Fig.\,\ref{radiomap} and \ref{radiomap1384}). 
Their locations and flux densities estimated using the \texttt{imsad} task are collected in 
Table\,\ref{newsources}. Only sources with significance above $10\,\sigma$ at both $5.5$\,GHz and at $1.384$\, GHz were considered.
Each of these sources has one or more potential counterparts in infrared and/or optical catalogues \R{\citep[see e.g.][]{2003yCat.2246....0C, 2010AJ....140.1868W, 2003PASP..115..953B, 2003AJ....125..984M}}. 
Two of them, J$1301-6306$ and J$1300-6311$, have X-ray counterparts in the \xmm\ catalog 
\citep{2016A&A...590A...1R}. The source J$1304-6258$ is actually visible in the MGPS-2 (see Fig.\,\ref{snr_counterpart} right) but is 
not listed in the catalogue probably because it is very difficult to separate its emission from the extended emission coincident with \snr.}

%% Most of the compact sources in the field 
%% were previously detected in the MGPS-2 \citep{2007MNRAS.382..382M} (Fig.\,ref{radiomap1384}).

%\footnote{http://skyview.gsfc.nasa.gov}

\section{Discussion}

%\subsection{Implication of lack of radio emission on the PWN \rr{properties}}

\rrr{The new radio observations reported in this paper were triggered by a hint of a signal ($\sim3\,\sigma$) 
from a feature coincident with the X-ray PWN in the analysis of archival data from the PMN survey 
at 4.85\,GHz \citep{2012A&A...548A..46H}. This feature is compatible to the \R{resolution} of the survey
of $\sim5^{\prime}$ \R{\citep{1993AJ....106.1095C}}. 
These deeper observations at 5.5\,GHz and 7.5\,GHz with \atca\ were optimised for the detection of the 
putative radio PWN with a size comparable to the size of the X-ray PWN ($\sim2^{\prime}$)
%% being sensitive to structures 
%% up to $\sim 2.7^{\prime}$ in extent at $5.5$\,GHz
while providing a resolution comparable to the one of \xmm. The assumption 
of the size was motivated by the size of the possible radio feature and by the hypothesis that the size of both radio 
and X-ray PWNe is constrained by the region of high magnetic field around the pulsar. Indeed, 3D magnetohydrodynamic simulations 
of the Crab Nebula \citep{2014MNRAS.438..278P} show that the magnetic field strength close to the termination 
shock is an order of magnitude higher than the average magnetic field strength in the rest of the PWN. 
Moreover, in the left-behind \RRR{relic} nebula the magnetic field is expected to be relaxed with the magnetic field strength
comparable to the interstellar medium (ISM) magnetic field of about $3\,\mu$G.
However, neither new observations at 
5.5/7.5\,GHz nor the analysis of the archival data at 1.384/2.368\,GHz show any evidence of extended emission 
coincident with \pulsar\ and/or the X-ray PWN.
\R{It should be noted that the largest scales to which observations at 5.5/7.5\,GHz are sensitive are slightly smaller than the
  size of X-ray PWN, $1.7^\prime$ at 5.5\,GHz and $1.3^\prime$ at 7.5\,GHz, and therefore we cannot rule out the detectebility
  of the putative radio PWN of the size of the X-ray PWN at these frequencies. }
\R{However, observations at 1.384/2.368\,GHz allow us to detect structures with an extension up to $\sim18^{\prime}$ (1.384\,GHz) and $\sim11^{\prime}$
  (2.368\,GHz). }
%% The largest structure which can be reliably imaged 
%% in this analysis is $\sim27^{\prime}$ at 1.384\,GHz.
Assuming the size of the X-ray PWN as reported in 
\citet{2012A&A...548A..46H} the upper limits on the radio flux 
\R{density} at
%$5.5$\,GHz \rr{and at $1.384$\,GHz} were estimate
$1.384$\,GHz was estimated
\rr{at the level of 3 time the RMS noise} 
%% using a bias-corrected bootstrap method 
%% \citep[see e.g.][]{bootstrap} based on the measured flux 
in the region of the X-ray PWN defined by 
a box (black box in Figs.\,\ref{radiomap} and \ref{radiomap1384}). 
%; emission from the pulsar was excluded in the upper limit calculation) 
%and performing 20.000 simulations. 
%% \rr{The values of the estimated upper limits 
%% %at $99\%$ confidence level is estimated to be 
%%   are $0.17$\,mJy at $5.5$\,GHz and $2.6$\,mJy at $1.384$\,GHz}.
\R{The upper limit on the flux density at $1.384$\,GHz is $2.6$\,mJy.}
%% \rr{The} upper limit at $5.5$\,GHz 
%% is much more constraining (by two orders of magnitude) than the one reported in \citet{2012A&A...548A..46H} for the insignificant radio 
%% feature of a similar size detected at $4.85$\,GHz. 

However, the size of the putative radio PWN might exceed the size of the largest angular structure resolved in the observations 
or even the size of the primary beam. GeV observaions \citep{2013ApJ...773...77A} exhibiting a large size 
of the PWN ($\sim0.9^\circ$) indicate that a large amount of relatively low energy relativistic electrons 
is spread out to large distances from the pulsar. If the size of the putative radio PWN is constrained 
by the existence of relativistic electrons, i.e. the magnetic field is strong enough for efficient 
synchrotron emission across the whole PWN, the radio PWN would be at least as large as the GeV 
PWN. The same electrons which emitt GeV \gammarays\ via inverse Compton scattering on ambient 
photon fields are also responsible for the generation of the radio emission via the synchrotron mechanism 
\citep[see e.g.][and references therein]{2013ApJ...773..139V}.}

%\subsection{The SNR candidate - a birth place of \pulsar?}

\RR{\subsection{\snr\ - an SNR?}}
\label{snr}

\begin{figure*}[ht!]
  \centering
  %\includegraphics[width=\textwidth]{allthree.pdf}

  %mgps2: scale -0.002 0.017
  % pmn: scale -0.05 0.5
  %1.384: scale -0.003 0.013

  \begin{subfigure}[b]{0.32\textwidth}
    \includegraphics[width=\textwidth]{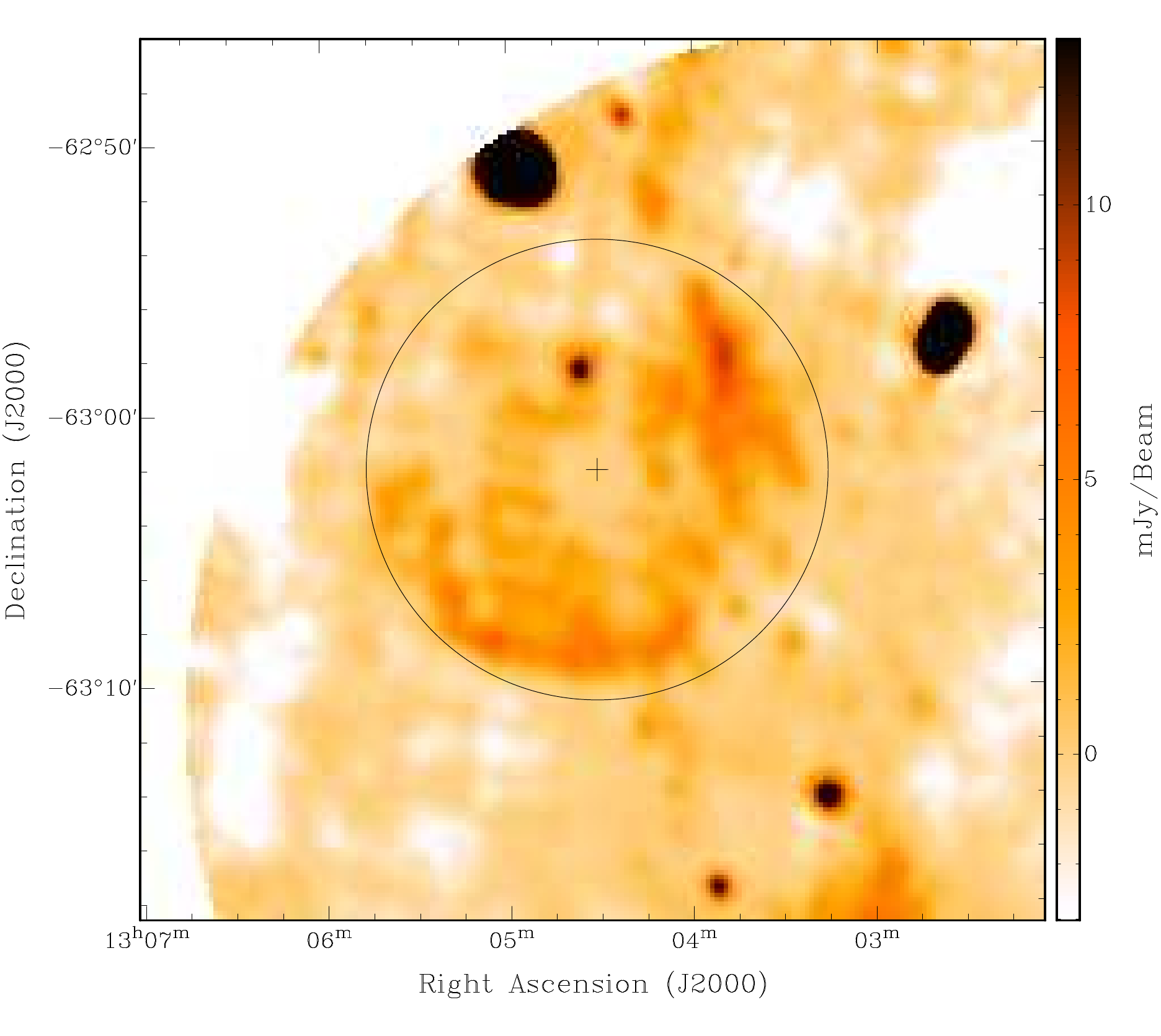}
  \end{subfigure}
  \begin{subfigure}[b]{0.32\textwidth}
    \includegraphics[width=\textwidth]{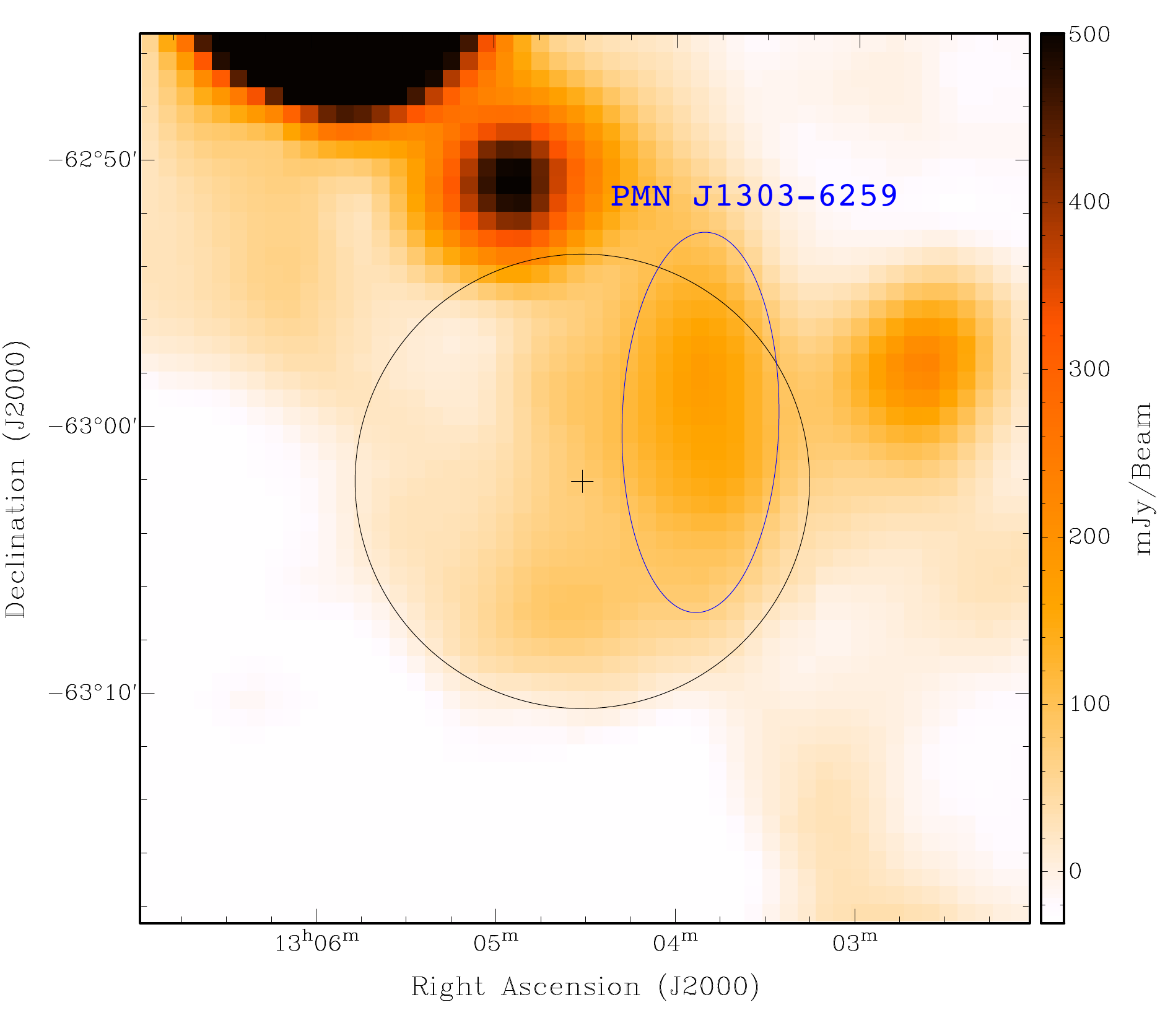}
  \end{subfigure}
  \begin{subfigure}[b]{0.32\textwidth}
    \includegraphics[width=\textwidth]{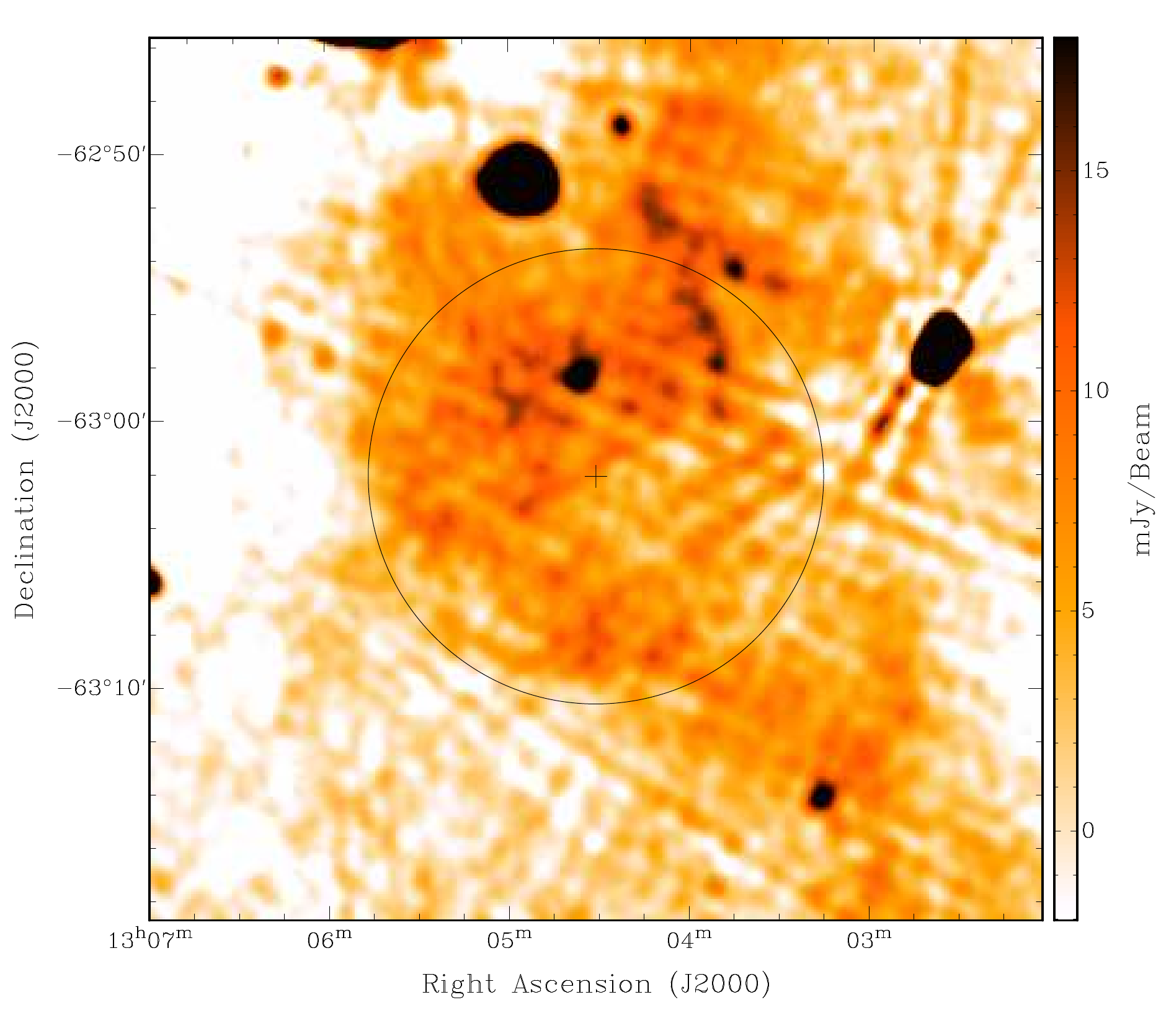}
  \end{subfigure}

 %mgps2: scale -0.002 0.018
  % pmn: scale -0.03 0.5
  %1.384: scale -0.003 0.013
  
  \caption{\rrr{Left: radio map of the \snr\ region at $1.384$\,GHz. Middle: radio map of the \snr\ region at $4.85$\,GHz obtained in the PMN survey 
      \citep{1993AJ....106.1095C}. The blue ellipse shows the extension of the PMN survey source \snrpmn\ \R{as presented in the catalogue \citep{1994ApJS...91..111W}. The resolution
      of this map is $\sim5^\prime$ \citep{1993AJ....106.1095C}}.
      Right: radio map of the \snr\ region at $843$\,MHz obtained in the MGPS-2 survey \citep{2007MNRAS.382..382M}}. \R{The size of the beam is $\sim45^{\prime\prime} \times 50^{\prime\prime}$.}
    \R{The black circle in all the panels indicates the extension
  of the shell-like structure and the black cross shows its centre. Middle and right images were obtained using the NASA SkyView online tool (\url{http://skyview.gsfc.nasa.gov}).}}
  \label{snr_counterpart}
\end{figure*}

\RR{The extended structure, \snr, detected to the east of the pulsar at 1.384 GHz exhibits a shell-like morphology.
  Such a morphology naturally suggests that this might be an SNR.} \rrr{
  %The SNR candidate
\RR{\snr} is coincident with \snrpmn, an extended source detected in the PMN survey at $4.85$\,GHz \R{\citep{1993AJ....106.1095C, 1994ApJS...91..111W}}. 
The emission from \snrpmn\ was fitted with a two-dimensional asymmetric Gaussian with major and minor axis widths of $\sigma_\mathrm{x} = 14.28^{\prime}$ and 
$\sigma_\mathrm{y} = 5.88^{\prime}$ with a position angle of $1.5^{\circ}$ measured eastwards from the north direction \R{\citep{1993AJ....105.1666G}}. The total flux \R{density} from the source is 
$235\pm14$\,mJy at $4.85$\,GHz \R{\citep{1994ApJS...91..111W}}. Although \snrpmn\ is coincident with the western part of the detected shell-like structure the map also reveals 
extended radio emission coincident with southern part of the shell (see Fig.\,\ref{snr_counterpart} middle). Overall, the $4.85$\,GHz map show a good agreement 
with the $1.384$\,GHz morphology of the SNR candidate. To estimate a spectral index of the western part of the SNR candidate we smoothed 
the $1.384$\,GHz map with a Gaussian with a width equal to the angular resolution of the PMN survey of $4.2^\prime$ and fitted the region 
of \snrpmn\ assuming the same extension of the source as obtained for \snrpmn\ using the \texttt{miriad} task \texttt{imfit}. The obtained 
flux \R{density} at $1.384$\,GHz is $398\pm3$\,mJy resulting in a spectral index of $\alpha = 0.42\pm0.05$\footnote{\R{Flux density, $S_\nu$,
scales with frequency, $\nu$, as $S_{\nu} \propto \nu^{-\alpha}$}}, which is 
in a perfect agreement with a range of values observed for SNRs \RR{from $\sim0.2$ to $\sim0.8$ with a peak at around $0.5$ \citep[][]{2011Ap&SS.336..257R, 2014BASI...42...47G}}
and very close to a canonical $\alpha = 0.5$ expected in the diffusive shock acceleration in the case of strong shocks with a compression
ratio of 4 \citep[see e.g.][and references therein]{2015A&ARv..23....3D}.}

\rrr{Neither the compact source catalogue nor the SNR catalogue of the second epoch Molonglo Galactic Plane 
Survey (MGPS-2) at $843$\,MHz \citep{2007MNRAS.382..382M, 2014PASA...31...42G} provide any counterpart for 
the SNR candidate. 
%% There are no apparent counterparts of the SNR candidate found in the the second epoch Molonglo Galactic Plane 
%% Survey (MGPS-2) at $843$\,MHz\citep{2007MNRAS.382..382M}. The search for the SNR candidates in the MGPS-2 image 
%% database also does not provide any counterpart with \snr\ \citep{2014PASA...31...42G}.
However, the MGPS-2 image 
of the \snr\ region at $843$\,MHz reveals a faint extended radio emission coincident with the SNR candidate, 
roughly following the $1.384$\,GHz contours in the west and south regions but also exhibiting emission towards north-east 
(Fig.\,\ref{snr_counterpart}, right). The peak flux \R{density} reaches $17.5$\,mJy/beam, well above the sensitivity of the 
survey of $2$\,mJy/beam. The image is strongly distorted by artefacts produced by nearby bright sources which makes it 
impossible to classify the morphology of the extended emission. This might be the reason why the source did not 
appear in the list of the SNR candidates detected in the MGPS-2 survey as the search criteria include a condition on a 
morphology which has to be shell-like or composite \citep{2014PASA...31...42G}.}

\rr{Infrared and optical} observations, e.g. the Two Micron All-Sky Survey (2MASS) 
in the H-band \citep[$1.65\,\mu$m;][]{2006AJ....131.1163S}, do not show any 
extended emission from the region of \rrr{\snr}. This \rrr{further} suggests 
that the radio emission from the \rrr{SNR candidate} is most probably 
non-thermal as expected for SNRs.

\RR{Available X-ray observations\footnote{\RR{All observations of the region of interest available at
    NASA's HEASARC archive (https://heasarc.gsfc.nasa.gov/docs/archive.html) were examined
  including \xmm, \swift, \rosat, \chandra, and \asca.}} of this region show no evidence
  of the large scale structures coincident with \snr. This, however, does not
  contradict the SNR hypothesis. The absence of the non-thermal X-ray emission can
  be explained by a \RRR{potentially} old age of the SNR candidate (see below). \RRR{Indeed,
  X-ray synchrotron emission from SNRs requires relatively high shock velocities of
  $\gtrsim2000$ km/s to accelerate electrons to energies high enough, and these shock
  velocities are \RRRR{believed to be associated with} young SNRs of $\sim1000$\,yr \citep[see e.g.][]{2012A&ARv..20...49V}.
  Old age \RRRR{might} also be a reason for the lack of the X-ray thermal emission. Slowing
  down of the shock to $<200$\,km/s in old SNRs of $\gtrsim10$\,kyr results in cooling
  of the post-shock region to temperatures lower than required for the X-ray emission
  \citep[see e.g.][]{2012A&ARv..20...49V}. Some old SNRs, however, feature thermal X-ray emission from their interior which might be due to, for example, interaction with dense cloudlets, which survive the forward shock crossing and slowly evaporate inside the remnant due to saturated thermal conduction \citep[see e.g.][]{1991ApJ...373..543W, 2011A&A...525A.154S}. Another possible reason for the the lack of the thermal emission might be \RRR{a} low density of the ambient medium around the remnant. How low should this density be is, however, model dependent and depends on whether ionization quilibrium is sustained. Non-equilibirum ionization models coupled to efficient particle acceleration shows that X-ray thermal emission would dominate over synchrotron emission only for relatively high densities of about $1$\,cm$^{-3}$ \citep{2007ApJ...661..879E}. The estimate of the density also depends on the distance to the remnant, which is not clear in this case. \RRRR{Finally, the lack of detection of X-ray emission from the source, both thermal and non-thermal, might be due to not sufficient sensitivity or exposure.}}

  At \gammaray\ energies most of the field of view is covered
  by \pwn\ and it is impossible to distinguish any emission associated with the
  possible SNR.}

\RR{It should be noted, however, that other possible interpretations of the detected
  extended emission cannot be firmly ruled out. Due to the faint emission the image of
  \snr\ appears to be pattchy making it difficult to firmly identify the shell-like morphology.
  The emission could also be contaminated by unrelated point sources and the image might be
  distorted by artefacts created by bright sources in the field of view. Since the shell-like
  morphology of the source is the strongest argument in favor of its SNR nature, we have to treat
  our conclusions with caution. The observed large scale structure might still be
  a composition of individual background sources and/or HII regions. 
}

\RR{\subsection{Possible birth place of \pulsar}}
\rrr{\RR{If \snr\ is indeed an SNR it} could be 
a birth place of the pulsar \pulsar.} 
\rrr{This hypothesis is supported by the direction of the \RR{X-ray} ``tail'' which roughly points to the position of the SNR candidate (see Fig.\,\ref{radiomap1384}).} 
The pulsar is located outside the shell-like structure, which means 
that in case the SNR candidate is indeed the birth place of \pulsar, the pulsar has already escaped the remnant and continues to 
propagate in the ambient medium. \rr{The larger size of the PWN does not contradict this hypothesis as the 
present time nebula could have formed after the pulsar escaped the SNR. While the pulsar is still inside the SNR, its 
nebula is strongly disrupted by the reverse shock of the remnant \citep[see e.g.][]{2001ApJ...563..806B, 2013A&A...551A.127V} 
and at the moment of interaction with the shell it is very small. Moving outside the remnant the pulsar builds up a new nebula which can become very large 
due to the proper motion of the pulsar, i.e. left behind electrons, and diffusion of electrons in the medium. 
Escaping the SNR, the pulsar should also damage the shell of the remnant.} 
%% This fact together with the large size of the GeV/TeV PWN (larger than the 
%% SNR candidate) suggests that there should be some evidence of distortion of the shell caused by the escape 
%% of the pulsar. 
\rrr{Although the emission in the direction of the pulsar is slightly fainter exhibiting a gap in the shell, 
there is no \RR{clear} evidence of distortion.} This can be naturally explained if the pulsar is not moving in the projected plane but its 
velocity has a considerable component perpendicular to the projected plane. In this case the distorted part 
of the shell is facing the observer and is thus not visible. The angular distance between the pulsar and 
\rr{the centre of} the SNR candidate of about \rr{$19^{\prime}$} corresponds to a projected distance of $36$\,pc 
assuming the distance to the pulsar of $6.6$\,kpc. This corresponds 
to a \rr{projected} pulsar velocity of \rrr{$V_{\mathrm{p}}\simeq 3100$\,km/s for the characteristic age of the pulsar of 11 ky \citep{2005AJ....129.1993M}}. \rrr{This velocity 
would make \pulsar\ the fastest known pulsar. The highest pulsar velocity detected so far is 
$\sim1600$\,km/s \citep{1998ApJ...505..315C}.} 
%%  \rr{This configuration, however, would mean that the 
%% pulsar velocity is considerably higher than its projection.} 
%% %% This configuration, however, requires a 
%% %% higher pulsar velocity depending on the angle $\phi$ between the line of sight and the pulsar proper motion 
%% %% direction 
%% %% \begin{equation}
%% %% V_{\mathrm{p}} = V_{0}/ \sin{\phi},
%% %% \end{equation}
%% %% where $V_{0} = 1000$\,km/s is the lower limit of the pulsar velocity calculated for $\phi = 90^{\circ}$.
%% The association of the pulsar with the SNR candidate would place \pulsar\ among the fastest known pulsars with 
%% highest detected velocity of $\sim1600$\,km/s \citep{1998ApJ...505..315C}.
\rr{However, there are indications that 
some pulsars might be much faster\rrr{, with velocities comparable to the estimate presented here for \pulsar}. 
The estimate of the kick velocity of the possible pulsar IGR\,J$11014-6103$ is 
$2400-2900$\,km/s \citep{2012ApJ...750L..39T}, but the source nature cannot be unambiguously proven yet as no pulsations 
were detected so far.} 
\rrr{Also, the real age of \pulsar\ might be higher than the characteristic age if the braking index is lower than $3$, 
  which is the case for 8 out of 9 pulsars for which the braking index has been measured reliably \citep{2015MNRAS.446..857L, 2016ApJ...819L..16A}.
  In this case the estimate of the pulsar velocity would be lower.}

\RR{
  Alterntively, the real age of the pulsar can be estimated as the age of its SNR.
  In case \snr\ is indeed an SNR and the birth place of the pulsar \pulsar, its size can
  provide an estimate for the pulsar age.
  %%   In case the extended structure with a shell-like morphology detected at 1.384 GHz is an SNR it 
%% could be the birth place of the pulsar \pulsar. 
Assuming a distance to the pulsar of $6.6$\,kpc \citep{2012A&A...548A..46H}, the angular size of the SNR candidate of about $16^{\prime}$ 
corresponds to 30\,pc in diameter. The Sedov solution \citep{1959sdmm.book.....S}, which describes 
the hydrodynamical expansion of an SNR in the adiabatic stage of evolution into the homogeneous medium, 
provides an estimate of the SNR age for a given size of the remnant
\begin{equation}
t_{\mathrm{age}} = 16 \left(\frac{E}{10^{51}\,[\mathrm{erg}]}\right)^{-1/2}\left(\frac{n_{\mathrm{ISM}}}{1\,[\mathrm{cm}^{-3}]}\right)^{1/2}\left(\frac{R}{15\,[\mathrm{pc}]}\right)^{5/2}\,[\mathrm{kyr}],
\end{equation}
where $E$ is the explosion energy, $n_{\mathrm{ISM}}$ is the number density of the interstellar 
medium and $R$ is the radius of the remnant. This value is somewhat higher than
%in a good agreement with
the characteristic age of the pulsar of 11 kyr \citep{2005AJ....129.1993M}, but does not contradict it if the
braking index is lower than 3 (see above). If we assume that the real age of the pulsar is 16 kyr then
the projected velocity would be $2100$\,km/s, which is still very high.
It should be noted however hat the SNR age estimate is dependent on the ambient medium density which is often
considerably lower than $1\,\mathrm{cm}^{-3}$ (even by \RRR{orders of} magnitude) and it is also very sensitive to
the estimated physical size of the SNR. Therefore, this estimate of age should be taken with caution.}

\RR{A pulsar with this high velocity will be moving supersonically in the interstellar medium creating a
  bow shock. A bow shock driven through the neutral gas can generate optical emission in the Balmer lines
  at the forward shock \citep{2001A&A...375.1032B, 2002A&A...393..629B} and such emission is already
  discovered for a few pulsars in H$_\alpha$. There is no evidence for such emission around \pulsar, which however does not
  necessarily contradict our hypothesis, since a lot of pulsars are believed to be moving supersonically and
  only for a few of them a bow-shock structure is detected at optical wavelengths. It should also be noted that
  H$_\alpha$ bowshocks and X-ray tails are rarely seen together \citep{2015SSRv..191..391K}.
  Inside the bow shock pulsar wind particles will be accelerated at the termination shock subsequently generating non-thermal emission
  synchrotron emission which can be detected in radio and X-rays \citep{2006ARA&A..44...17G}. Although no
  extended radio emission associated with \pulsar\ was detected (which can be simply because of the potentially
  large size of the radio nebula), the X-ray nebula does exhibit a bow-shock
  morphology with a tail pointing in the direction of \snr\ strongly suggesting that the pulsar is moving
  with high velocity.}

\section{Summary}
\label{summary}

\atca\ observations of the \pwn\ \rr{region} at $5.5$\,GHz and $7.5$\,GHz do not 
reveal any significant extended emission associated with \pulsar. Archival 
$1.384$\,GHz and $2.368$\,GHz data also do not show any evidence of a radio 
counterpart of \pwn. Non-detection of this evolved PWN at radio wavelengths 
suggests that either the putative radio PWN is comparable to the size of the 
GeV counterpart and, thus, larger than the \rrr{largest reliably imaged structure 
and even the primary beam} of radio observations or that 
the magnetic field is rather low, which is in agreement with the evolved 
PWN identification as the magnetic field is expected to decrease with time 
in PWNe. \RRR{The comparison of the X-ray emission to the TeV emission implies the
average magnetic field in the PWN of $\sim 0.2- 2\,\mu$G \citep{2012A&A...548A..46H}, which however does not exclude the possibility of enhanced magnetic field around the pulsar.}

Archival $1.384$\,GHz observations reveal a detection of a \RRRR{extended} structure \R{centred} at the angular 
\R{distance} of \rr{$19^\prime$} from the pulsar. This \RRRR{extended} structure might be an 
SNR and a potential birth place of the pulsar. If this is the case \R{then}
the projected velocity of the pulsar \R{would be} \rrr{$3100$\,km/s \RR{assuming the characteristic age
    of the pulsar. This would make \pulsar\ the fastest known pulsar. However, the uncertainty of the
    true age of the pulsar can significantly change this estimate.}}

\acknowledgements

We would like to thank the anonymous referee for valuable comments which strongly improved the paper.

\R{We would like to acknowledge the help of Marek Jamrozy (Astronomical Observatory of the Jagiellonian University of Krak\'ow, Poland) and Michael Bietenholz (Hartebeesthoek Radio Observatory, South Africa; York University, Toronto, Canada) on understanding certain aspects of the ATCA analysis performed in the paper.}

The Australia Telescope Compact Array is part of the Australia Telescope National Facility which is funded by the Australian Government for operation as a National Facility managed by CSIRO. 

This paper includes archived data obtained through the Australia Telescope Online Archive (\url{http://atoa.atnf.csiro.au}).



\bibliographystyle{aa} 
%\bibliography{references_full}
\bibliography{j1303_atca_aanda_final.bbl}


%% \begin{thebibliography}{00}


%% \bibitem[ ()]{}

%% \end{thebibliography}
\end{document}